\documentclass[letterpaper,twocolumn,10pt]{article}
\usepackage{usenix2019_v3}
\usepackage{enumitem}
\usepackage{tikz}
\usepackage{amsmath}
\usepackage{float}
\usepackage{enumitem}
\usepackage{placeins,adjustbox}
\usepackage{threeparttable}
\usepackage{xcolor}
\usepackage{comment,booktabs}
\usepackage{xspace}

\newcommand*\circled[1]{\tikz[baseline=(char.base)]{
            \node[shape=circle,draw,inner sep=2pt] (char) {#1};}}

\newcommand{\tool}{PolyScope\xspace}

\newcommand{\myparagraph}[1]{\vspace{0.25em}\noindent\textbf{#1:}}
\let\paragraph=\myparagraph

\newcommand{\trent}[1]{\textcolor{orange}{\textbf{TRENT:} #1}}
\newcommand{\eddy}[1]{\textcolor{brown}{\textbf{EDDY:} #1}}

\begin{document}


\date{}

\title{\Large \bf PolyScope: Multi-Policy Access Control Analysis to Triage Android Systems}


\author{
{\rm Yu-Tsung Lee}\\
Penn State University
\and
{\rm William Enck}\\
North Carolina State University
\and
{\rm Haining Chen}\\
Google
\and
{\rm Hayawardh Vijayakumar}\\
Samsung Research
\and
{\rm Ninghui Li}\\
Purdue University
\and
{\rm Daimeng Wang}, {\rm Zhiyun Qian}\\
University of California, Riverside
\and
{\rm Giuseppe Petracca}~\thanks{Giuseppe Petracca's work on this paper was performed when he was a graduate student at Penn State.}\\
Lyft
\and
{\rm Trent Jaeger}\\
Penn State University
} 

\maketitle

\begin{abstract}
Android's filesystem access control provides a foundation for Android system integrity.
Android utilizes a combination of mandatory (e.g., SEAndroid) and discretionary (e.g., UNIX permissions) access control, both to protect the Android platform from Android/OEM services and to protect Android/OEM services from third-party apps.  However, OEMs often create vulnerabilities when they introduce market-differentiating features because they err when re-configuring this complex combination of Android policies.  In this paper, we propose the \tool tool to triage the combination of Android filesystem access control policies to vet releases for vulnerabilities.  The \tool approach leverages two main insights: (1) adversaries may exploit the coarse granularity of mandatory policies and the flexibility of discretionary policies to increase the permissions available to launch attacks, which we call {\em permission expansion}, and (2) system configurations may limit the ways adversaries may use their permissions to launch attacks, motivating computation of {\em attack operations}. 
We apply \tool to three Google and five OEM Android releases to compute the attack operations accurately to vet these releases for vulnerabilities, finding that permission expansion increases the permissions available to launch attacks, sometimes by more than 10X, but a significant fraction of these permissions (about 15-20\%) are not convertible into attack operations.  
From \tool's results, we describe two previously unknown vulnerabilities and show
how \tool helps OEMs triage the complex combination of access control policies down to attack operations worthy of testing.
\end{abstract}
\vspace{-0.1in}

\section{Introduction}
\vspace{-0.05in}
Android has become the most dominant mobile OS platform worldwide, deployed by a large number of vendors across a wide variety of form factors, including phones, tablets, and wearables~\cite{statcounter}.  With Android's increased integration into people's daily lives, Android needs to provide sufficient and appropriate assurances of platform integrity.  Additionally, vendors must be able to extend the Android platform to support their custom functionality and yet maintain such assurances to their customers.  Android's implementation of filesystem access control is one of the most important defenses for providing such assurances.

While Android adopted advanced access control methods aggressively, such as SEAndroid mandatory access control~\cite{smalley2013security} (MAC),  
in combination with various traditional access control methods, such as UNIX discretionary access control (DAC), 
many vulnerabilities in filesystem access have still been reported.  
In one recent case reported by Checkpoint~\cite{MITD}, an untrusted application abused write permission to Android's external storage to replace other applications' library files with fake library files before the victim application installs them, in what is known as a {\em file squatting attack}.  In another instance reported by researchers from IOActive \cite{IOACTIVE}, a vulnerability in the DownloadProvider allowed untrusted applications to read/write unauthorized files by providing a maliciously crafted URI that causes a victim's DownloadProvider to access an untrusted symbolic link that redirects the victim to the targeted files, which is called a {\em link traversal attack}. This vulnerability could have serious effects since some over-the-air update files, including for some privileged applications, are also downloaded by the DownloadProvider.  

Researchers have proposed automated policy analysis to detect misconfigurations that may expose vulnerabilities in complex access control policies~\cite{jaeger02sacmat,setools}, but application of these methods to Android does not address: (1) how those policies may be altered by adversaries and (2) how to detect which operations adversaries may actually be able to employ in attacks on those misconfigurations.
The emergence of Android with its rich permission system and its subsequent adoption of the SEAndroid mandatory access control motivated the development of policy analysis methods for Android systems~\cite{enck09ccs,wae+17,Wang2015,aafer15ccs}.  However, each of these initial approaches only considered a single type of access control policy (e.g., either SEAndroid or Android permissions).  
Recently, work has been proposed to compute the information flows of combined MAC and DAC policies~\cite{chen17acsac}, including Linux capabilities as well~\cite{BigMAC}.  However, these techniques miss some attacks, such as the recent Checkpoint and IOActive attacks referenced above, because they lack methods to detect how adversaries may broaden their ability to launch attacks by manipulating the flexibility inherent in the UNIX DAC and Android permission systems.  In addition, these techniques may identify many spurious threats because they do not determine whether adversaries can really launch attacks for the threats found.  

In this paper, we aim to develop a method to triage Android filesystem access control policies by: (1) identifying
the resources applications are authorized to use that are modifiable by their adversaries, accounting for policy manipulations, and (2) determining the attack operations on those resources available to adversaries to enable testing the victims for vulnerabilities.  Like several prior works in access control policy analysis~\cite{jaeger02sacmat,setools,jaeger03usenix,chen09ndss}, our method starts by identifying filesystem resources at risk by computing the integrity violations authorized.  An {\em integrity violation} occurs when an access control policy authorizes a lower-integrity subject (adversary) to modify a resource used by a higher-integrity subject (victim).  However, an integrity violation may not imply a vulnerability because: (1) victims may not actually use the impacted resource and/or (2) adversaries may not be able to exploit the victim's use of the resource.  Predicting the resources a program may use in advance is a difficult challenge, so we focus instead on computing whether and how an adversary could attempt to exploit a victim's use.  For each integrity violation, we compute the ways that adversaries may launch attacks, which we call {\em attack operations}.

To compute attack operations for Android systems comprehensively and accurately, we leverage the insight that adversaries may exploit flexibility in Android access control to expand the permissions available to themselves and/or victims to launch attack operations, which we call {\em permission expansion}.  DAC protection systems allow resource owners to grant permissions to their resources arbitrarily, making prediction of whether an unsafe permission may be granted intractable in general~\cite{hru76}.  Adversaries can leverage such flexibility to grant victims permissions to lure them to resources to which adversaries can launch attack operations.  In addition, Android systems convert many Android permissions that adversaries may obtain into DAC permissions, leading to further risks. 
While SEAndroid MAC policies bound such permissions, these MAC policies are sufficiently coarse that changes in DAC permissions may reveal many new attack operations within the MAC restrictions.  

We build a static access control analysis tool called \tool that computes attack operations on filesystem resources from Android access control policies and host/program configurations.  \tool is targeted for use by Android system vendors (OEMs) who often extend base Android systems with a variety of vendor-specific services and applications to customize their products with value-added functionality.  OEMs must then configure Android access control to protect their value-added functionality and the system's trusted computing base,
OEMs face challenges in determining whether the resultant access control configurations reveal new vulnerabilities.  In Section~\ref{subsec:case} we describe vulnerabilities caused by misconfiguration of access control policies for two such value-added functionalities.  
\tool enables OEMs to compute the attack operations that third-party apps may launch against their functionality to identify actions that may lead to vulnerabilities or at least require vulnerability testing.  In addition, OEMs can also use \tool to compute the attack operations that their vendor-specific services and applications may be able to launch, should they be compromised, against the Android services to evaluate protection of their system's trusted computing base.  

Our evaluation in Section~\ref{sec:eval} demonstrates that \tool has several benefits over prior analysis approaches and is practical to use.  First, in a study of seven freshly installed Android releases\footnote{We examine an eighth system in some experiments, which has a significantly greater number of pre-installed apps.}, three Google Android versions and four OEM Android versions, we find that permission expansion increases the number of integrity violations significantly, from (122\% to 1550\%) across versions.  However, between 14\% and 21\% of those integrity violations cannot be transformed into attack operations by the filesystem and/or program configurations.  Second, \tool finds that OEM releases have a significantly greater number of attack operations than the Google releases.  Using these attack operations uncovered by \tool, we find two new vulnerabilities in three OEM releases through manual analysis.  One of these new vulnerabilities requires permission expansion to exploit, demonstrating the power of \tool.  Finally, we implement an analysis method in \tool that enables parallel validation of potential integrity violations, resulting in significant performance improvements for the studied systems, ranging from 70\% to 84\% improvement across releases.  This suggests that OEMs may benefit from applying \tool incrementally to identify attack operations as they extend their systems with new value-added features. 

This paper makes the following contributions:

\vspace{-0.07in}
\begin{itemize}
    \item \textit{We propose the \tool analysis tool for Android filesystem access control.}
    \tool composes Android's access control policies and relevant system configurations to compute the attack operations available to adversaries, accounting for permission expansion.
    
    \item \textit{We use \tool to triage three Google and five OEM Android releases.}
     We find a significantly greater number of attack operations for OEM's Android releases, indicating that OEMs may greatly benefit from \tool as they customize their Android-based products. 
    
    \item \textit{We identify two new vulnerabilities in OEM Android releases.}
    Using \tool results, we identify vulnerabilities in: (1) the Thememanager used by Xiaomi and Huawei and (2) Samsung's Resestreason logging.
    
\end{itemize}
\vspace{-0.07in}

The remainder of this paper is as follows.
Section~\ref{sec:motivation} motivates the need for more effective access control analysis.  
Section~\ref{sec:overview} provides an overview of the proposed \tool approach.
Section~\ref{sec:threat} defines our threat model.
Sections~\ref{sec:design} and~\ref{sec:impl} describe the design and implementation of \tool.
Section~\ref{sec:eval} performs a variety of experiments to show how \tool triages access control policies in Android releases.
Section~\ref{sec:discussion} describes current limitations and how they may be addressed.
Section~\ref{sec:relwork} examines differences from related work.
Section~\ref{sec:conc} concludes the paper.

\vspace{-0.1in}
\section{Motivation}
\vspace{-0.07in}
\label{sec:motivation}

In this section, we motivate the goals of our work.  We first present an example of the challenge of detecting the attack operations that may lead to vulnerabilities in Android systems in Section~\ref{subsec:example}.  After outlining current approaches to access control policy analysis in Section~\ref{subsec:ivs}, we describe their key limitations in Section~\ref{subsec:limits}.

\vspace{-0.1in}
\subsection{An Example Vulnerability}
\label{subsec:example}
\vspace{-0.05in}

A recent vulnerability discovered in Android services using the DownloadProvider allows untrusted apps to gain access to privileged files~\cite{IOACTIVE}.
The DownloadProvider enables services to retrieve files on behalf of apps by a URI specifying the location of a file.
An untrusted app may lure a service's DownloadProvider into using a maliciously crafted URI that resolves to a symbolic link created by the untrusted app.  Through this symbolic link, the untrusted app can access any file to which the service is authorized, which may include some privileged files.  This is an example of a {\em link traversal attack}. 

\begin{figure}[t]
\includegraphics[width=2.5in]{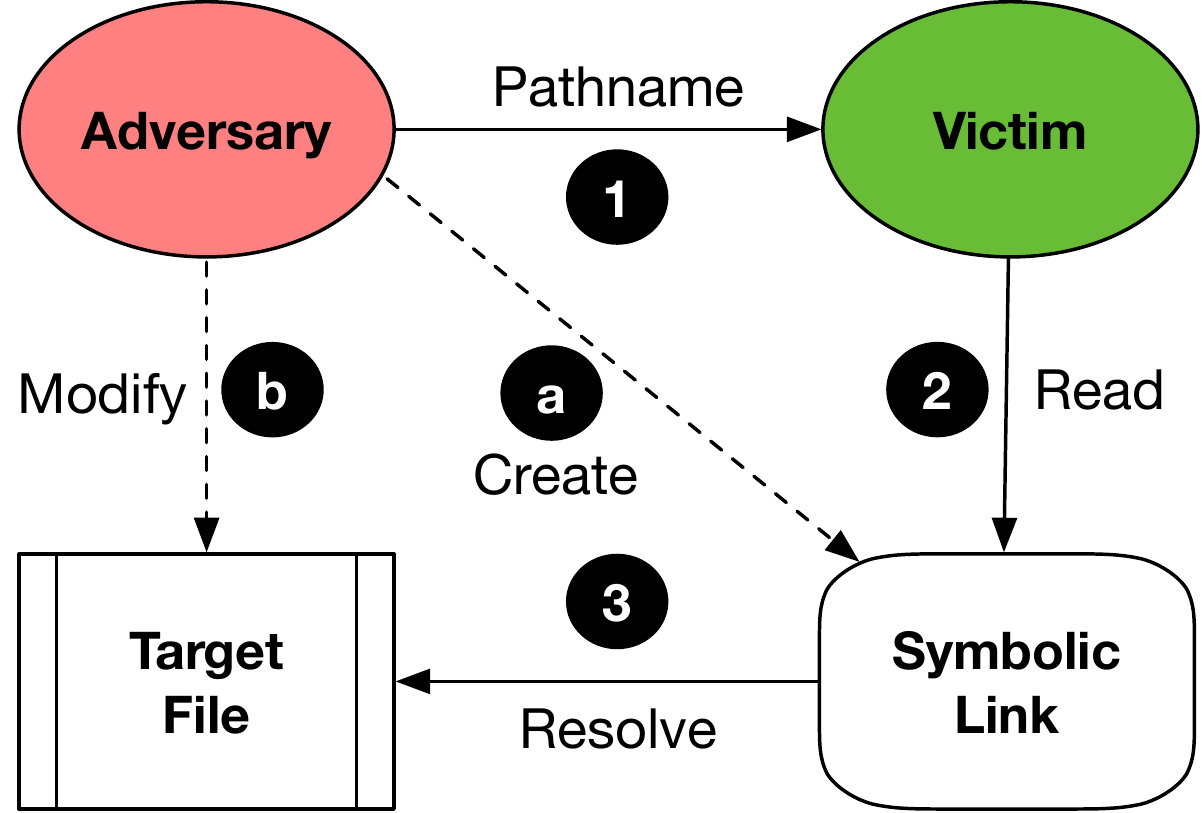}
\centering
\caption{{\bf DownloadProvider Vulnerability}: (1) Adversary provides pathname to victim (as URI) to (2) lure the victim to an adversary-created symbolic link (a) that (3) the victim resolves to the target file enabling the adversary to modify the file indirectly through the victim (b).}
\label{fig:example}
\vspace{-0.1in}
\end{figure}


Figure~\ref{fig:example} shows exploitation of the vulnerability.  The adversary sends a request URI (Pathname in Figure~\ref{fig:example}) to the victim (service running DownloadProvider) {\footnotesize\circled{1}} that directs the victim to a symbolic link created by the adversary {\footnotesize\circled{a}}.  When the victim uses its read permission to the symbolic link  {\footnotesize\circled{2}}, the operating system resolves the link {\footnotesize\circled{3}} to return access to the target file.  This vulnerability may enable the adversary to leak and/or modify the target file {\footnotesize\circled{b}} to which the adversary normally lacks access.

This vulnerability occurred because adversaries of the service have the permission to create a symbolic link in a directory to which the service running DownloadProvider also has access.  Android access control aims to limit the exposure of services to directories modifiable by third-party apps and other adversaries, but sometimes functional requirements demand such permissions be made available.  In addition, Android systems allow apps to extend their own permissions by obtaining Android permissions and to grant services permission to access app directories, expanding the directories at risk.  

\vspace{-0.1in}
\subsection{Access Control Policy Analysis}
\label{subsec:ivs}
\vspace{-0.05in}

Seminal work in access control policy analysis~\cite{jaeger02sacmat,setools} proposed computing authorized information flows among subjects and objects from a system's access control policies.  An access control policy authorizes an {\em information flow from a subject to an object} if the policy allows that subject to perform an operation that modifies the object, called a {\em write-like operation}, and authorizes an {\em information flow from an object to a subject} if the policy allows that subject to perform an operation that uses the object's data, called a {\em read-like operation} (e.g., read or execute).  Some operations may be both read-like and write-like, enabling information flow in both directions.

However, modern Android systems have hundreds of thousands of access control rules, so there are many, many authorized information flows.  Researchers then explored methods to find the information flows associated with potential security problems. Some access control analyses focus on identifying secrecy problems~\cite{chen17acsac,enck09ccs,wae+17,Wang2015,aafer15ccs} and others on integrity problems~\cite{jaeger03usenix,chen09ndss}.  In the example above, this vulnerability permits attacks on process integrity by controlling the file retrieved by the victim process, whereby the compromised process may be directed to leak or modify files on behalf of the adversary.  To detect integrity problems, access control analyses are inspired by integrity models, such Biba integrity~\cite{b77}, to detect information flows from adversaries to victims.  Such information flows are called {\em integrity violations}, which are defined more formally as a tuple of resource, adversary, and victim, where the access control policy authorizes an information flow from the adversary to the resource (i.e., the adversary is authorized to perform a write-like operation on the resource) and authorizes an information flow from the resource to the victim (i.e., the victim is authorized to perform a read-like operation on the resource). 

\begin{figure*}[t]
\includegraphics[width=6.5in]{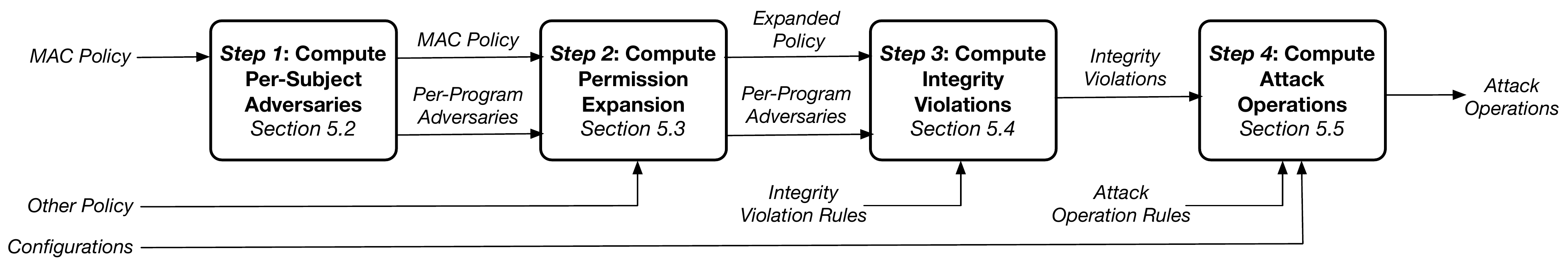}
\centering
\caption{{\bf \tool Logical Flow:} \tool computes per-subject adversaries, permission expansion by those adversaries, the integrity violations adversaries are authorized, and the attack operations adversaries may perform to attack victims.}
\label{fig:flow}
\vspace{-0.1in}
\end{figure*}

\vspace{-0.1in}
\subsection{Limitations of Current Techniques}
\label{subsec:limits}
\vspace{-0.05in}

Access control policy analyses attempt to solve three main problems to help identify vulnerabilities, but current approaches suffer from key limitations on each problem.

The first problem in using access control policy analysis is to {\bf identify the adversaries who may benefit from exploitation of each subject}.  Previous research often identifies untrusted apps\footnote{Includes apps assigned to the SEAndroid domains {\tt untrusted\_app} and {\tt isolated\_app}.  Information on SEAndroid domains appears in Section~\ref{subsec:background}.} as adversaries, and assumed that system services and OEM's value-added apps and services were trusted.  However, as OEMs push more functionality into their own Android distributions, they deploy a variety of new and modified apps and services whose trustworthiness may vary.  A recent study~\cite{gamba2019analysis} reveals that some OEM pre-installed code has a lack of end-to-end quality control and might even leverage code from third parties,   
resulting in back doors and other vulnerabilities~\cite{BH2017}.  By ignoring OEM apps and services, we risk missing attacks that utilize them as stepping stones to exploit Android system services.  However, we must be careful not to overapproximate adversaries, which leads to false positives.

The second problem is to {\bf determine the permissions adversaries control to create integrity violations that may enable attacks}.  Recent access control analysis methods that reason about multiple access control policies~\cite{chen17acsac,BigMAC} do not account for how an adversary may exploit flexibility in those policy models to manipulate permissions to increase the integrity violations they may attempt to exploit.  In Android systems, both the DAC access control model and the Android permission system allow untrusted parties to modify policies.  While in theory MAC policies could be used instead to govern access to prevent such problems, MAC policies tend to be more complex to configure and are unforgiving if a needed permission is not granted.  Thus, we have found that OEMs often configure DAC policies to protect their value-added apps and services and allow users to choose some authorizations using Android permissions.  Researchers have previously identified problems caused by DAC policy flexibility, ranging from the inability to predict whether an unauthorized permission could be granted~\cite{hru76} to the inability to prevent compromised subjects from propagating attacks~\cite{flask-inevitability}.  

The third problem is to {\bf compute the operations that an adversary may be authorized to employ to launch attacks}, which we call {\em attack operations}.  Once we know that a true adversary may be authorized permissions that enable integrity violations, a question is how an adversary may utilize such permissions to launch attacks.  While integrity violations are a necessary precondition to launch an attack, other configurations may prevent some operations useful for attacks.  Android systems provide a variety of ad hoc configurations that could prevent some attack operations.  For example, Android employs filesystem configurations to prevent symbolic links from being created in external storage directories, which can prevent link traversal attacks like the example above.  In addition,  Android systems have also introduced a specialized FileProvider class that requires that clients open files for their servers, which also can prevent link traversal attacks.  However, such ad hoc configurations are only employed sporadically, and recently suggested configuration operations also provide incomplete defense (see Section~\ref{subsec:scoped}).

\vspace{-0.1in}
\section{\tool Overview}
\vspace{-0.07in}
\label{sec:overview}


In this paper, we present a new Android access control analysis tool, called \tool, that computes the set of authorized attack operations for an Android system while overcoming the prior limitations described above.  Figure~\ref{fig:flow} shows \tool's approach.  In Step 1, \tool identifies the adversaries for each subject using definitions of mutual trust validated against an approach that computes worst-case, as described in Section~\ref{subsec:adversary}.  In Step 2, \tool determines the permissions adversaries control by modeling how adversaries may expand permissions available to themselves and their victims by exploiting the flexibility in Android access control policies, as described in Section~\ref{subsec:expand}.  In Step 3, \tool uses these expanded permissions to compute integrity violations based on {\em integrity violation rules} defined in Section~\ref{subsec:ivdef}.  In Step 4, \tool uses these integrity violations to compute the types of attack operations possible using {\em attack operation rules} defined in Section~\ref{subsec:operations}.  We identify the specific types of integrity violations and attack operations we consider in this paper in Section~\ref{sec:threat}.  

\tool computes integrity violations and attack operations to triage Android releases for  vulnerabilities induced by access control policies.  Integrity violations identify the filesystem resources that victims are authorized to access that their adversaries are authorized to modify (see Section~\ref{subsec:ivs}).  Attack operations specify the types of operations that adversaries are capable of performing in modifying filesystem resources to launch attacks.  Using this information, an analyst can detect the victim operations that risk compromise by using a filesystem resource in an integrity violation and the types of attack operations for which should be tested.  Detecting whether the victim is vulnerable requires testing the victim against instances of these attack operations.   In Section~\ref{sec:impl}, we describe some ways to detect victim operations, and we find two new vulnerabilities from subsequent testing in Section~\ref{subsec:case}.

\vspace{-0.1in}
\section{Threat Model}
\label{sec:threat}
\vspace{-0.07in}

In this paper, adversaries may modify any part of the filesystem to which they are authorized.  Adversaries may modify a part of the filesystem that a potential victim may be authorized to use and/or may send arbitrary requests (e.g., IPCs) to victim processes to which they are authorized to communicate.  These scenarios result in integrity violations (IVs), which we aim to compute accurately.  We assume that an adversary will try to exploit any integrity violation using any attack operations they are allowed to run to exploit the victim's use of the associated files.  

\begin{figure}[t]
\includegraphics[width=2.5in]{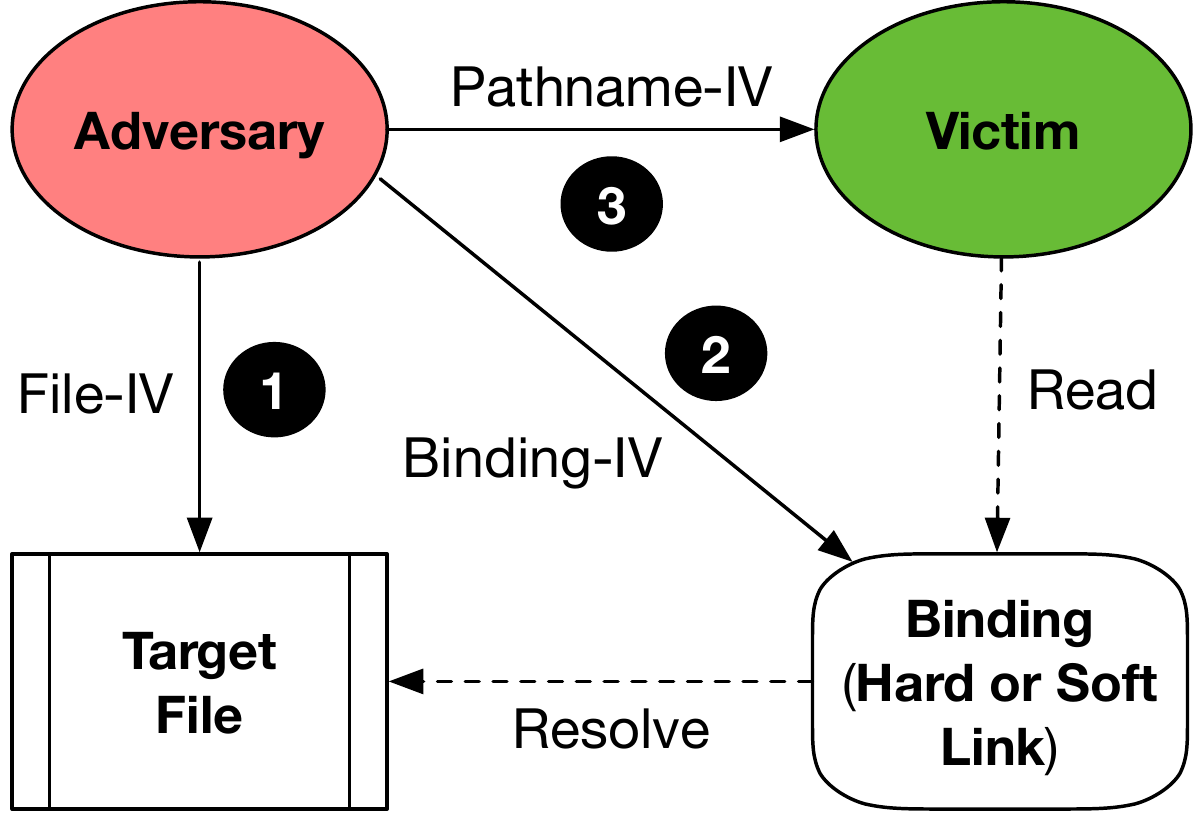}
\centering
\caption{{\bf Integrity Violation (IV) Classes}: (1) {\em File-IVs} grant adversaries direct access to modify files that victims use; (2) {\em Binding-IVs} grant adversaries the ability to modify name resolution of file names; and (3) {\em Pathname-IVs} enable adversaries to lure victims to the part of the filesystem they can modify.}
\label{fig:classes}
\vspace{-0.1in}
\end{figure}


In this paper, we aim to triage systems for three classes of integrity violations that authorize adversaries to perform attack operations on filesystem access, covering a wide variety of vulnerabilities including confused deputy~\cite{confused-deputy} and time-of-check-to-time-of-use (TOCTTOU) vulnerabilities~\cite{mcphee74,bishop-dilger}.  Related to Figure~\ref{fig:example}, we show these integrity violation classes in Figure~\ref{fig:classes}.  First, {\em file-IVs} allow adversaries to modify target files that are authorized to victims directly {\footnotesize\circled{1}}, possibly leading victims to unexpected use adversary-controlled data.  Second, {\em binding-IVs} enable adversaries to redirect victims to target files during name resolution {\footnotesize\circled{2}}, causing victims to operate on files chosen by adversaries.  Third, {\em pathname-IVs} enable adversaries to lure victims to an adversary-controlled part of the filesystem using an adversary-supplied pathname {\footnotesize\circled{3}}, which is the integrity violation exploited in the example vulnerability of Section~\ref{subsec:example}. 


For each integrity violation found, we assume that an adversary may attempt any possible attack operation.  In general, attack operations enable adversaries to provide malicious input to victims by getting them to use an adversary-controlled file or binding to enable the adversary to choose the input to the victim, whose basic approaches date to the 1970s~\cite{mcphee74}.  File-IV attack operations simply modify the resource awaiting use by the victim.  Binding-IV attack operations direct the victim to a resource chosen by the adversary, using link traversal or file squatting attacks.  A {\em link traversal} attack aims to exploit a victim as a confused deputy~\cite{confused-deputy} by directing the victim to access a resource to which the adversary is not authorized.  A {\em file squatting} attack aims to exploit a victim by planting an adversary-controlled resource at a location where the victim expects a protected file, like attacks on file-IVs.  Both types of attack operations may exploit TOCTTOU flaws~\cite{bishop-dilger}, if present, in the victim.  Pathname-IVs attack operations lure a victim who processes pathnames (e.g., URLs) to an adversary-controlled binding to exploit a link traversal.  


In developing \tool, we assume trust in some components of Android systems.  First, we assume that the Android operating system operates correctly, including enforcement of its access control policies and system configurations correctly.  For example, we trust the Android operating system to satisfy the reference monitor concept~\cite{anderson72}.  We note that the Android operating system includes the Linux operating system and a variety of system services. 
Our assumptions about trust among such services is determined using Android specifications, as described in Section~\ref{subsec:adversary}.

\vspace{-0.1in}
\section{\tool Design}
\label{sec:design}
\vspace{-0.07in}


In this section, we examine the design challenges in computing attack operations for Android systems.  In particular, after providing some brief background information, we focus on four key steps outlined in the \tool overview in Section~\ref{sec:overview}.  

\vspace{-0.1in}
\subsection{Design Background}
\label{subsec:background}
\vspace{-0.05in}

In this section, we aim to provide background on the various access control techniques employed by Android systems necessary to understand the \tool design.  Android deploys SEAndroid mandatory access control (MAC), Linux/UNIX discretionary access control (DAC), the Android permission system, and Linux capabilities to control access to filesystem resources directly or indirectly.  Linux capabilities have no tangible impact on contributing attack operations on recent Android versions, so we do not discuss them further.  Using the remaining models, we define \tool's interpretation of {\em subjects} and {\em objects} applied in policy analysis.

\paragraph{SEAndroid MAC}
SEAndroid is a port of SELinux mandatory access control system~\cite{selinux} with additional support for Android mechanisms, such as Binder IPC. SEAndroid supports three access control models: Type Enforcement (TE), Role-Based Access Control (RBAC), and Multi-Level Security (MLS). All the models are {\em mandatory access control models} (MAC) in that they are defined by the system and are not intended to be modified by users or their programs.  Out of these three, the TE model acts as the primary enforcement model to protect the integrity of the Android system's trusted computing base processes. MLS is deployed mainly to separate apps from one another. On the other hand, RBAC does not receive much use currently on Android, so we will not describe it further.  

The SEAndroid TE policy\footnote{Note that in this paper we sometimes refer to the "MAC TE" policy, which is the same as the SEAndroid TE policy.} defines authorization rules in terms of security labels~\cite{BoKa85}, where a subject can perform an operation on an object if there is a rule authorizing the subject's security label to perform the operation on object's security label. The SEAndroid MLS policy enables subjects and objects to be associated with categories~\cite{blp76}, where subjects can only perform operations on objects when the subject is authorized for the object's category.

\paragraph{Linux/UNIX DAC}
Android systems also utilize traditional UNIX discretionary access control (DAC) as provided by the Linux system on which Android is based.  UNIX DAC associates files with a UID for the file owner and a GID for the file group.  Processes are also associated with a process UID and GID, but a process may additionally belong to a set of supplementary groups.  A process can perform an operation on a file if: (1) the file's UID is the same as the process's UID and the file owner is authorized to perform that operation; (2) the file's GID matches one of the process's groups (i.e., process's GID or supplementary) and the file group is authorized to perform that operation; or (3) any process UID (i.e., others) is authorized to perform that operation for that file.  Importantly, UNIX DAC allows a process to modify file permissions when the process's UID is the same as the file's owner UID. 

Rather than associating UID's with individual users, as is traditional, Android associates UIDs with individual services and apps.  Thus, services and apps "own" a set of files (i.e., with the app's UID as the file owner UID) for which they may modify permissions.  Thus, malicious apps can change the permissions for files they own, which is important for luring victims to create pathname-IVs.

\paragraph{Android Permission System}
Android permissions are employed to control access to app and service data.  Android data/service providers enforce most Android permissions, but some Android permissions are mapped to DAC supplementary groups, which are assigned to apps when the associated Android permission is granted.  Thus, Android permissions may add DAC supplementary groups to app processes, granting them additional filesystem permissions.

Note that each Android permission has an associated \emph{protection level} that is used to determine whether or not an application may be granted that permission.
Over time, the permission granting policy has become more complex~\cite{zg16}.
Currently, permissions with the "normal" protection level are automatically granted to applications.
However, permissions with the "dangerous" protection level (e.g., guarding sensitive personal data such as GPS) require additional runtime authorization from the user.
Permissions with the "signature" protection level can only be granted to applications signed by the same developer key that was used to define the permission.
The signature protection level is primarily used to restrict access to functionality that only system applications should access.
Finally, there are several other flags that provide ad hoc restrictions, e.g., a "privileged" flag allows privileged, OEM-bundled applications to acquire associated permissions.  

\paragraph{Mapping MAC and DAC Policies}
To reason about access control for the combined DAC and MAC policies, \tool needs to determine how to map MAC policies in the form of TE security labels and MLS categories to DAC policies in the form of UIDs and groups.  Fortunately, Android makes such determination straightforward.  Files and directories are explicitly assigned both MAC and DAC information directly, so there is no possibility of ambiguity.  For processes, the mapping between MAC and DAC information is indirect.  Android assigns the same MAC TE security label, MAC MLS category, DAC UID, and DAC groups\footnote{The complete set of supplementary groups assigned to a program's processes depend on the Android permissions obtained for the program.  We define the assumption we use for \tool in Section~\ref{subsec:expand}.} to each program when it is run, as identified by Chen et al.~\cite{chen17acsac}.  Thus, we collect the MAC-DAC mapping for processes by running programs.  For all apps and services we have run, this relationship has held, but this implies that we can only perform policy analysis for apps and services installed on the release (i.e., that we can run).  We use this information to define subjects and objects for \tool analysis as follows.  

\begin{itemize}
    \item {\bf Subjects}: Each process is associated with a {\em subject} defined by its MAC TE label, MLS category set, DAC UID, and a set of DAC groups (GID and supplemental groups).  There may be many processes associated with one subject.
    \item {\bf Objects}: Each resource is associated with an {\em object} defined by its MAC TE label, MLS category set, DAC UID/GID, and mode bits (i.e., owner, group, others permissions).  There may be many files/directories associated with one object.
\end{itemize}

\tool reasons about access control policies in terms of subjects and objects, rather than individual processes and resources, as the definitions of subjects and objects form equivalence classes with respect to adversaries.  All processes of the same subject share the same adversaries, and all resource associated with the same object are modifiable by the same adversaries.  Thus, we express \tool results in terms of subjects and objects in Section~\ref{sec:eval}.

\vspace{-0.1in}
\subsection{Compute Per-Subject Adversaries}
\label{subsec:adversary}
\vspace{-0.05in}

\begin{table*}[ht]
\centering
\begin{threeparttable}%
    \caption{Google's Process Privilege Levels~\cite{PPRIV_LEVEL}}
    \small
\begin{tabular}{c|l}
 \textbf{Process Level$^1$} &\textbf{Level Membership Requirements}\\  \midrule
Root Process (T5)    & Process running with UID root (e.g., MAC labels {\tt kernel} and {\tt init})       \\
System Process (T4)   & Process running with UID system (e.g., MAC label {\tt system server})       \\
Service Process  (T3)   & AOSP core service providers (e.g., MAC labels {\tt bluetooth} and {\tt mediaserver})      \\
Trusted Application Process (T2)    & AOSP default and vendor apps (e.g., MAC labels {\tt platform\_app} and {\tt priv\_app})      \\
Untrusted Application Process (T1)     & Third-party applications (e.g., MAC label {\tt untrusted\_app})        \\
Isolated Process  (T0) &  Processes that are expected to receive adversarial inputs (e.g., MAC label {\tt webview})         \\
\end{tabular}
\begin{tablenotes}\scriptsize
    \item[1] Listing types of processes based on their privilege level, from high to low with root processes being most privileged (T5) and isolated processes being the least privileged (T0).  We group T0 and T1 together calling the resultant level T1 in the evaluation in Section~\ref{sec:eval}.
\end{tablenotes}
\label{table:levels}
\end{threeparttable}%
\end{table*}

A challenge is to identify the adversaries for each subject.  Researchers often identify subjects adversaries based untrusted sources (e.g., Chen et al.~\cite{chen17acsac} used third-party apps as adversaries) or based on their role in the system (e.g., Jaeger et al.~\cite{jaeger03usenix} said only core system services could be trusted).  A problem is that these approaches are one-dimensional and ad hoc.  Since they are ad hoc, the likelihood of missing possible attack sources or identifying trusted sources as false adversaries is greater. Since they are one-dimensional (i.e., either focusing on trust or not), there is no basis to determine whether adversaries are missing or misclassified.

We propose to develop a method for computing adversaries that considers both the best-case and worst-case trust to derive and validate per-subject adversary sets.  For the worst case, we leverage the conservative {\em integrity wall} approach of Vijayakumar {\em et al.}~\cite{integrity-wall}.  The integrity wall approach makes the insight that only the subjects that can trivially compromise a subject's program must be trusted, thus computing a minimal trusted computing base (TCB) (i.e., fewest subjects trusted) of subjects for each program.  For the best case, we utilize the process privilege levels defined by Google~\cite{PPRIV_LEVEL}, which groups subjects in classes by whether they should be mutually trustworthy.  By examining trust from two directions, we can perform validation on whether the combination is consistent.  While this approach enables just one type of limited validation, we are not aware of any prior work validating adversary sets.

The integrity wall method computes per-subject TCBs by detecting whether either one of two requirements are met: (1) that subject must trust any other subjects that are authorized to modify files that the subject may execute (i.e., its executable and library files) and (2) that subject must trust any other subjects that are authorized to modify kernel resources. We compute worst-case per-subject TCBs from MAC TE policies. 

On the other hand, Android specifies "privilege levels"~\cite{PPRIV_LEVEL} that describe which subjects should mutually trust one another, implying a best-case TCB. Google defines six privilege levels in an Android system~\cite{PPRIV_LEVEL}, which we group into five levels for evaluation in Section~\ref{sec:eval}: (T5) root processes; (T4) system processes; (T3) service processes; (T2) trusted application processes; (T1) untrusted application processes and isolated process.  Isolated apps and untrusted apps are separated into distinct privilege levels by Google, but in this paper, we do not consider attacks on untrusted apps by the lower privileged isolated apps.  Table~\ref{table:levels} lists these privilege levels based on their UID and MAC labels.  

Using Google's privilege levels, we assume a subject trusts all of the subjects at its level or higher.  For example, untrusted apps trust other untrusted apps and any subjects at higher privilege levels, such as the Android system services (e.g., system server).  Trusting subjects at higher privilege levels is accepted because such subjects provide functionality that the subjects at lower privilege levels depend upon.  However, assuming untrusted apps may be mutually trusting may be harder to accept, but we are not looking for attacks between untrusted apps in this paper.  Resolving how to identify adversaries among untrusted apps, such as determining whether mutual trust for all is appropriate, is future work.

To produce an accurate adversary set, we validate consistency between the best-case and worst-case trust sets to derive an adversary set that is consistent with both trust sets.  Specifically, \tool validates whether the worst-case trust set for each subject is a subset of that subject's best-case trust set.  If so, then there does not exist an adversary of any subject relative to its best-case trust set (i.e., fewest adversaries) that is not also an adversary relative to that subject's worst-case trust set (i.e., maximal adversaries).   An inconsistency implies that the Android privilege levels are missing a fundamental trust requirement to prevent trivially compromising the subject.  

\vspace{-0.15in}
\subsection{Compute Permission Expansion}
\label{subsec:expand}
\vspace{-0.05in}

A key difficulty for OEMs is predicting which resources may be accessible to adversaries and victims to derive attack operations accurately.  A problem is that while MAC policies are essentially fixed (i.e., between software updates), DAC permissions may be modified by adversaries to increase the attack operations that they could execute.   We identify two ways that adversaries may modify permission assignments on Android systems: (1) adversaries may obtain Android permissions that augment their own DAC permissions and (2) adversaries may delegate DAC permissions for objects that they own to potential victims.  For some Android permissions, adversaries gain new DAC permissions to access additional resources that may enable attacks.  By delegating DAC permissions to objects they own, adversaries may lure potential victims to resources that adversaries control.  

\paragraph{Adversary Permission Expansion}
In Android systems, some Android permissions are implemented using DAC groups.  As described above, a process is associated with a single UID and GID, but also an arbitrarily large set of supplementary groups that enable further "group" permissions.  Thus, when a user grants an Android permission associated with one or more DAC groups to an app, there is a direct expansion of that app's permissions in terms of its DAC permissions.  Since the MAC policies are generally lenient in Android systems, these new DAC permissions may grant privileges that enable attacks.  For \tool, we assume that subjects can obtain all of their "normal" Android permissions and are granted all of their "dangerous" permissions by users for analysis, as described in the previous section.  One of the vulnerability case studies we highlight in Section~\ref{subsec:case} exploits adversary permission expansion.  

\paragraph{Victim Permission Expansion}
Researchers have long known that allowing untrusted subjects (i.e., adversaries in our case) to administer DAC permissions for objects that they own can present difficulties in predicting possible permission assignments.  Researchers proved that the {\em safety problem} for a typical DAC protection system, i.e., an access matrix for subjects and objects where objects and permissions may be added in a single command, is undecidable in the general case~\cite{hru76}.  The safety problem asks whether there exists a general algorithm to predict whether a particular unsafe permission will ever be granted to a particular subject given an initial state.  As a result, researchers explored alternative DAC models for which the safety problem could be solved, such as the take-grant model~\cite{take-grant}, the typed access matrix~\cite{tam}, and policy constraints~\cite{tidswell00ccs}. 

Using the ability to manage DAC permissions to objects they own, adversaries can grant permissions to their resources to victims, expanding the set of resources that victims may be lured to use.  In many cases, victims have MAC permissions that grant them access to adversary directories, but vendors use DAC permissions to block access.  However, since adversaries own these directories, they can simply grant the removed permissions to potential victims themselves.  

We can predict the permissions available to victims despite the undecidability of the safety problem because we do not have to solve the general problem.  First, we find the resources to which the victim has MAC permissions to access and the adversaries own.  Only those resources are worth granting DAC permissions, as the victim can only access if it has MAC permissions as well.  Thus, the number of objects on which DAC permissions may be usefully granted to victims is finite.  

\vspace{-0.12in}
\subsection{Compute Integrity Violations}
\label{subsec:ivdef}
\vspace{-0.05in}

In this section, we show how to compute integrity violations for file-IVs, binding-IVs, and pathname-IVs defined in Section~\ref{sec:threat}.  Recall from Section~\ref{subsec:ivs} that integrity violations are a tuple of resource, adversary, and victim,  where the adversary is authorized to modify the resource and the victim is authorized to use (e.g., read, write, or execute) the resource.  


\paragraph{Computing File Integrity Violations}
A file vulnerability may be possible if a subject uses (read, write, or execute) a file that can be modified by an adversary.  In practice, many subjects read files their adversaries may write ({\tt read-IVs}) with adequate defenses, but risks are greater if the subject executes ({\tt exec-IVs}) or also modifies such files ({\tt write-IVs}).  For exec-IVs, executing input from an adversary enables an adversary to control a victim's executable code.  For write-IVs, if a subject writes to a file its adversaries also may write, then adversaries may be able to undo or replace valid content.

\vspace{-0.07in}
\begin{scriptsize}
\begin{verbatim}
{read|write|exec}(file, victim) &&  // victim has access to file,
adv-expand(adversary) &&            // but adversary-expanded perms      
write(file, adversary)              // enables to modify file
\end{verbatim}
\vspace{-0.1in}
$\rightarrow$
\vspace{-0.08in}
\begin{verbatim}
{read|write|exec}-IV(file, victim, adversary)
\end{verbatim}
\end{scriptsize}
\vspace{-0.07in}

This rule determines whether the victim is authorized by the combination of access control policies for reading, writing, or executing files, using the {\tt \{read|write|exec\}} predicate.  The rule accounts for the adversary's expansion of their own permissions, as indicated by the predicate {\tt adv-expand}.  If the adversary also has write permission to the file ({\tt write} predicate), then the associated integrity violation is created. 

\paragraph{Computing Binding Integrity Violations}
A binding vulnerability is possible if a subject may use a binding in resolving a file name that adversaries can modify.  

\vspace{-0.07in}
\begin{scriptsize}
\begin{verbatim}
use(binding, victim) &&            // victim can use binding,
adv-expand(adversary) &&  // but adversary-expanded perms
write(binding, adversary)          // enable to modify binding
\end{verbatim}
\vspace{-0.1in}
$\rightarrow$
\vspace{-0.08in}
\begin{verbatim}
binding-IV(binding, victim, adversary)
\end{verbatim}
\end{scriptsize}
\vspace{-0.07in}

This rule parallels the rule for file-IVs, except that this rule applies to a victim having the permission to use a binding in name resolution ({\tt use} predicate). 

\paragraph{Computing Pathname Integrity Violations}
Pathname integrity violations are a type of binding integrity violation that are possible when a subject uses input from adversary to determine the bindings to use in name resolution.  These integrity violations place additional requirements on the adversary to be authorized to communicate with the victim.  In addition, adversaries can apply victim permission expansion to lure victims to a larger set of bindings the adversaries are authorized to modify.   

\vspace{-0.07in}
\begin{scriptsize}
\begin{verbatim}
write(ipc, adversary, victim) &&       // may send IPCs to victim
vic-expand(adversary, victim) &&       // and expand victim perms
binding-IV(binding, victim, adversary) // to lure victim to binding
\end{verbatim}
\vspace{-0.1in}
$\rightarrow$
\vspace{-0.08in}
\begin{verbatim}
pathname-IV(binding, victim, adversary)
\end{verbatim}
\end{scriptsize}
\vspace{-0.07in}

Adversaries must be granted write privilege to communicate to the victim, as defined in the {\tt write} predicate.  Android services may use Binder IPCs, and we further limit {\tt write} to use IPCs that communicate URLs for Android services.  The adversary can use victim expansion to increase the set of bindings the victim is authorized to use by {\tt vic-expand}.  If that binding meets the requirements of a binding-IV (see above), then a pathname-IV is also possible for this victim.

\vspace{-0.12in}
\subsection{Compute Attack Operations}
\label{subsec:operations}
\vspace{-0.03in}

We define how \tool computes attack operations from the integrity violations computed in the last section and the relevant system configurations.  We identify four types of attack operations that an adversary could use to exploit the three types of integrity violations: (1) file modification for file IVs; (2) file squatting for binding-IVs; (3) link traversal for binding-IVs; and (4) luring traversal for pathname-IVs.  

\paragraph{File Modification Attacks}
For read/write/exec IVs, the attack operation is to modify the objects involved in each IV.  However, Android uses some read-only filesystems, so not all files to which adversaries have write privilege are really modifiable.  Thus, the rule for {\em file modification} operations additionally checks whether the file is in a writable filesystem.

\vspace{-0.05in}
\begin{scriptsize}
\begin{verbatim}
{read|write|exec}-IV(file, victim, adversary) &&
fs-writable(file)      // file's filesystem is writable
\end{verbatim}
\vspace{-0.1in}
$\rightarrow$
\vspace{-0.08in}
\begin{verbatim}
file-mod(file, victim, adversary)
\end{verbatim}
\end{scriptsize}
\vspace{-0.05in}

\paragraph{File Squatting Attack}
In a file squatting attack, adversaries plant files that they expect that the victim will access.  The adversary grants access to the victim to allow the victim to use the adversary-controlled file.  This attack operation gives the adversary control of the content of a file that the victim will use.  To perform a file squatting attack operation, the adversary must really be able to write to the directory to plant the file.  Thus, the rule for {\em file squatting} operations is essentially the same as for file modification, but applies to binding-IVs.  

\vspace{-0.05in}
\begin{scriptsize}
\begin{verbatim}
binding-IV(binding, victim, adversary) &&
fs-writable(binding)     // binding's filesystem is writable
\end{verbatim}
\vspace{-0.1in}
$\rightarrow$
\vspace{-0.08in}
\begin{verbatim}
file-squat(binding, victim, adversary)
\end{verbatim}
\end{scriptsize}
\vspace{-0.05in}

In this rule, we assume that the adversary predict the filenames used by the victim.  In the future, we will explore extending the rule to account for that capability.

\paragraph{Link Traversal}
A link traversal attack is implemented by planting a symbolic link at a binding modifiable by the adversary, as described in Section~\ref{subsec:example}.  However, Android also uses some filesystem configurations that prohibit symbolic links, so not all bindings to which adversaries have write privilege and are in writable filesystems allow the creation of the symbolic links necessary to perform link traversals.  Thus, the rule for {\em link traversal} operations extends the rule for file squatting to account for this additional requirement.  

\vspace{-0.05in}
\begin{scriptsize}
\begin{verbatim}
binding-IV(binding, victim, adversary) &&
symlink(binding) &&      // binding's filesystem allows symlinks
fs-writable(binding)     // binding's filesystem is writable
\end{verbatim}
\vspace{-0.1in}
$\rightarrow$
\vspace{-0.08in}
\begin{verbatim}
link-traversal(binding, victim, adversary)
\end{verbatim}
\end{scriptsize}
\vspace{-0.05in}

   
\paragraph{Luring Traversal}
An adversary may lure a victim to a binding controlled by the adversary to launch an attack operation.  However, Android systems also offer a defense to prevent luring the victim to open files on behalf of adversaries in the form of the FileProvider class.  Specifically, the FileProvider class requires that clients open files themselves and provide the FileProvider with the resultant file descriptor.  Since the clients open the file, they perform any name resolution, so the potential victim is no longer prone to pathname vulnerabilities.  Thus, the rule for {\em luring traversal} operations extends the rule for link traversal for pathname-IVs by requiring the absence of any FileProvider class usage.  OEMs still have many services and privileged apps that do not employ the FileProvider class, so there remain several opportunities for pathname-IVs to be attacked.

\vspace{-0.05in}
\begin{scriptsize}
\begin{verbatim}
pathname-IV(binding, victim, adversary) &&
symlink(binding) &&        // binding's filesystem allows symlinks
fs-writable(binding) &&    // binding's filesystem is writable
no-file-provider(victim)   // victim does not employ fileprovider
\end{verbatim}
\vspace{-0.1in}
$\rightarrow$
\vspace{-0.08in}
\begin{verbatim}
luring-traversal(binding, file, victim, adversary)
\end{verbatim}
\end{scriptsize}
\vspace{-0.05in}


While it is possible that the victim has implemented an extra defense in Android middleware (e.g., Customized Android Permission) to prevent IPCs, we do not yet account for these defenses, as these permissions are ad hoc.  Including these defenses is future work.

\vspace{-0.12in}
\section{Implementation}
\label{sec:impl}
\vspace{-0.07in}

The \tool tool is implemented fully in Python in about 1500 SLOC and is compatible with Android version 5.0 and above.  
After {\em Data Collection} gathers access control policies, \tool implements the logical flow shown in Figure~\ref{fig:flow} in a slightly  different manner described below.  First, \tool computes integrity violations in steps one to three in Figure~\ref{fig:flow}, but only for the SEAndroid TE policy, which we call {\em TE IV Computation}.  Next, \tool computes whether the TE integrity violations are authorized by the remaining Android access control policies by re-running steps two and three in Figure~\ref{fig:flow}, but only for resources associated with the TE IVs, which we call {\em TE IV Validation}.  This separation enables us to parallelize the validation step, which has a significant performance impact, see Section~\ref{subsec:perf}.  Finally, \tool leverages the validated IVs to {\em Compute Attack Operations}.  Below, we discuss these major phases of the implementation, and how we use the results in {\em Testing for Vulnerabilities}.

\paragraph{Data Collection} We implemented multiple data collection scripts that collect access control data for the subjects and objects from an Android phone.  The methods are relatively straightforward for accessible files and processes, detailed in Appendix~\ref{subsec:collect}.  However, we are not authorized to access all files, particularly those owned by root, so we run these scripts on rooted phones.  Recent work by Hernandez et al.~\cite{BigMAC} is able to extract MAC policy and part of DAC configuration from Android firmware images without rooting devices. However, it has difficulty obtaining files located in some directories like {\tt /data}. As shown in Table 1 of their paper~\cite{BigMAC}, about 75\% of the files' DAC configuration in {\tt /data} cannot be retrieved, which we extract with our scripts.

\paragraph{TE IV Computation} To compute per-subject adversaries in Step 1 of Figure~\ref{fig:flow}, \tool leverages the integrity wall~\cite{integrity-wall} and Android privilege levels~\cite{PPRIV_LEVEL}, as described in Section~\ref{subsec:adversary}.  We follow the procedure defined in the integrity wall paper for Linux~\cite{integrity-wall}, except we add objects upon which Android kernel integrity depends (e.g., {\tt rootfs} and {\tt selinuxfs}) to the set of kernel objects. 
Since the SEAndroid TE policy is immutable (i.e., modulo system upgrades), Step 2 of Figure~\ref{fig:flow} is not required.  In Step 3, \tool computes the integrity violations authorized by the TE policy, as specified in Section~\ref{subsec:ivdef}. 

\begin{table*}[t]
\centering
\begin{threeparttable}[b]
    \caption{Summary of Impact of \tool Analyses}
    \label{tab:sum_table}
     \footnotesize
\begin{tabular}{@{}c|ccc|cccc@{}}
                                 & \multicolumn{3}{c|}{\textbf{Cross Version IVs (Google)}} & \multicolumn{4}{c}{\textbf{Cross OEM IVs}} \\
                                 & \textbf{Nexus 5x 7.0} & \textbf{Nexus 5x 8.0} & \textbf{Pixel3a 9.0} & \textbf{Mate9 8.0} & \textbf{Mate9 9.0}  & \textbf{Mix2 9.0} & \textbf{Galaxy S8 9.0} \\ \midrule
\textbf{Authorized Data Flows}$^1$         & 204,241               & 166,027           & 156,315             & 240,916              & 860,508                    & 289,238            & 259,992                \\
\textbf{IVs for \tool Adversaries}$^2$     & 167                   & 80                & 69                  & 223                  & 166                        & 192             & 628                  \\
\textbf{\tool IVs after Expansion}$^3$     & 372                   & 478               & 1,139               & 1,682                & 1,566                      & 2,304           & 4,377                \\
\textbf{\tool IVs with Operations}$^4$     & 297                   & 387               & 927                 & 1,331                & 1,327                      & 1,979           & 3,736                \\
\textbf{Average Operations per IV}$^5$     & 1.18                  & 1.10              & 1.04                & 1.62                 & 1.37                       & 1.08            & 1.36
\end{tabular}%
\begin{tablenotes}\scriptsize
    \item[] Unit is the relation \{Subject, Object\}, where subjects and objects are defined in Section~\ref{subsec:background}
\item[1] Objects authorized for use by Subjects 
\item[2] Authorized Data Flows where Object is modifiable by at least one \tool per-victim adversary, see Section~\ref{subsec:adversary}
\item[3] \tool Integrity Violations (sum for all types) as defined in Section~\ref{subsec:ivdef}
\item[4] \tool Integrity Violations in at least one Attack Operation, see Section~\ref{subsec:operations} 
\item[5] Sum of Attack Operations (in Table~\ref{tab:AOC}) per \tool IVs with Operation
\end{tablenotes}
\end{threeparttable}%
\end{table*}

\paragraph{TE IV Validation} After the TE IVs are identified, \tool validates whether these TE IVs are also authorized for the combination of remaining Android access control  policies: UNIX DAC, SEAndroid MLS, and Android permissions.  Step 1 of Figure~\ref{fig:flow} is not rerun.  In Step 2, \tool converts Android permissions to authorized DAC subgroups for adversary expansion and identifies the objects owned by each subject for victim expansion, as described in Section~\ref{subsec:expand}. In Step 3, each TE IV is evaluated to determine whether SEAndroid MLS and DAC policies also authorize the victim and adversary of each IV the permissions necessary to justify an integrity violation.  As mentioned above, the set of TE IVs can be partitioned to validate them in parallel.  

\paragraph{Compute Attack Operations} \tool then computes the attack operations for the IVs using the filesystem and program configurations as described in Section~\ref{subsec:operations}. 
\tool collects the relevant filesystem configurations by parsing the associated mount options.  
\tool collects the relevant program configurations (i.e., whether the victim includes a recommended defense, the FileProvider class) by reverse engineering the application's {\tt apk} package to detect the presence of the FileProvider class. 
We validated the ability or inability to perform attack operations and found no discrepancies.


\paragraph{Testing for Vulnerabilities}  The ultimate goal is to determine whether the victim is vulnerable to any  of the attack operations.  However, a key challenge is to determine whether and when a victim may actually access a resource associated with an attack operation.  Just because a potential victim may be authorized to use a resource, does not mean it ever uses that resource.  Even if a potential victim may use a resource associated with an attack operation, we need to determine the conditions when such an access is performed.  Thus, detecting vulnerabilities often requires runtime testing. 

The major challenge here is how to properly drive the victim subjects' programs so we can observe all file usage, akin to fuzz testing.  Developing a fuzz testing approach for the file system is outside the scope of this paper, so we simply drive programs with available tools: (1) Android Exerciser Monkey; (2) Compatibility Testing Suite (CTS); and (3) Chizpurfle ~\cite{chizpurfle}. We use the Android Exerciser Monkey and CTS to emulate normal phone usage, and Chizpurfle to drive Android system services.  With the accurate attack surface identified by \tool, we are able to find vulnerabilities described in Section~\ref{subsec:case}. 
We discuss how to employ runtime systematically in the future in Section~\ref{subsec:our_limits}.

\vspace{-0.1in}
\section{Evaluation}
\label{sec:eval}
\vspace{-0.07in}

Table~\ref{tab:sum_table} summarizes the highlights of our evaluation for seven fresh installs of OEM Android releases, demonstrating the importance of computing per-victim adversaries, permission expansion, and attack operations. Table~\ref{tab:sum_table} shows the relative effort to vet Android releases for vulnerabilities using the output of prior analyses~\cite{BigMAC,chen17acsac} ({\em Authorized Data Flows}), output of a past analyses~\cite{jaeger03usenix,integrity-wall} using \tool's method for computing adversaries ({\em IVs for \tool Adversaries}), and two new analyses performed by \tool ({\em \tool IVs after Expansion} and {\em \tool IVs with Operations}) that provide a more accurate accounting of the threats victims may face.  The counts are shown in terms of subject-object pairs, as {\em subjects} and {\em objects} are defined in Section~\ref{subsec:background}.  For data flows, we sum of the objects that each subject is authorized to use (i.e., in a {\em read-like operation}, see Section~\ref{subsec:ivs}).  For integrity violations (IVs), we only count the data flows to objects that another subject classified as an adversary is authorized to modify (i.e., in a {\em write-like operation}, see Section~\ref{subsec:ivs}).  

The first row of Table~\ref{tab:sum_table} lists the number of {\em authorized data flows} allowed by Android access control policies, showing that analyses that only compute data flows~\cite{BigMAC,chen17acsac} leave OEMs with hundreds of thousands of cases to assess to detect vulnerabilities in Android releases.  The second row in Table~\ref{tab:sum_table} shows that the number of data flows to consider can be reduced significantly by only considering those that cause integrity violations.   The particular way \tool computes adversaries per-victim (see Section~\ref{subsec:adversary}) for finding the {\em IVs for \tool adversaries} results in a reduction of the number of data flows involved in integrity violations by at least two orders of magnitude.  

Additionally, \tool provides two new analysis steps to detect threats more accurately.  First, the third row of Table~\ref{tab:sum_table} shows the number of {\em \tool IVs after expansion}, which shows the counts for IVs found using the rules defined in Section~\ref{subsec:ivdef}.  In several cases, the number of integrity violations increases significantly after accounting for expansion, in some cases by more than 10X.  Second, the fourth row of Table~\ref{tab:sum_table} shows that the number of {\em \tool IVs with operations} based on the rules in Section~\ref{subsec:operations} may be significantly reduced (14-21\% across these releases) because no attack operation may be possible for some IVs given the filesystem and/or victim subjects' program configurations.  The average number of attack operations possible for each integrity violation with at least one attack operation is shown in the fifth row, indicating the effort to test each release for vulnerabilities in terms of the types of operations that must be tested.   

In the remainder of the evaluation, we examine how the \tool implementation (see Section~\ref{sec:impl}) impacts the analysts' efforts to vet their releases for vulnerabilities in Sections~\ref{subsec:te-ivs} to~\ref{subsec:iv2ao}, we assess the distribution of IVs across privilege levels in Section~\ref{subsec:cross}, we describe how we found two types of previously unknown vulnerabilities using \tool output in Section~\ref{subsec:case}, and we measure \tool's analysis performance in Section~\ref{subsec:perf}.

\vspace{-0.1in}
\subsection{TE IV Computation}
\label{subsec:te-ivs}
\vspace{-0.05in}

\begin{table*}[t]
\centering
\begin{threeparttable}[b]
    \caption{Integrity Violations across Vendor Releases}
    \label{tab:attack_surface}
     \footnotesize
\begin{tabular}{@{}c|ccc|ccccc@{}}
                                 & \multicolumn{3}{c|}{\textbf{Cross Version IVs (Google)}} & \multicolumn{5}{c}{\textbf{Cross OEM IVs}} \\
                                 & \textbf{Nexus 5x 7.0} & \textbf{Nexus 5x 8.0} & \textbf{Pixel3a 9.0} & \textbf{Mate9 8.0} & \textbf{Mate9 9.0} & \textbf{Mix2 8.0}$^1$ & \textbf{Mix2 9.0} & \textbf{Galaxy S8 9.0} \\ \midrule
\textbf{MAC TE allow rules}$^2$         & 64,830                & 133,545               & 191,556              & 250,220             & 276,181            & 273,295           & 282,650           & 498,941                \\
\textbf{TE Write-IVs}$^4$           & 468               & 411                     & 1,130                & 2,067              & 1,958            & 1,657             & 2,197             & 1,787                \\
\textbf{TE Read-IVs}$^4$            & 1,410             & 2,373                   & 4,296                & 8,922              & 9,890            & 8,370             & 8,423             & 10,912                \\ 
\textbf{TE Binding-IVs}$^3$         & 495               & 4,38                    & 693                  & 1,504              & 1,233            & 1,400             & 1,174             & 1,881                  \\ \hline
\textbf{Valid Write-IVs}$^4$          & 120               & 19                      & 56                   & 400                & 236              & 232               & 216               & 469                    \\
\textbf{Valid Read-IVs}$^4$           & 194               & 80                      & 85                   & 679                & 437              & 531               & 749               & 953                    \\
\textbf{Valid Binding-IVs}$^3$        & 52                & 22                      & 32                   & 217                & 159              & 248               & 154               & 550                   \\
\textbf{Valid Pathname-IVs}$^4$       & 178               & 398                     & 1,054                & 1,003              & 1,129            & 1,186             & 1,555             & 3,424
\end{tabular}%
\begin{tablenotes}\scriptsize
    \item[] TE implies only having permission in SEAndroid TE
    \item[1] This phone has significantly more files perhaps related to a higher number of pre-installed apps
    \item[2] Unit: number of rules
    \item[3] Unit: IVs (victim, object) for directory objects only
    \item[4] Unit: IVs (victim, object) for file objects only
\end{tablenotes}
\end{threeparttable}%
\end{table*}

{\em How many integrity violations are found when using the SEAndroid MAC TE policy alone in TE IV Computation?}
\tool's implementation computes IVs initially using only the SEAndroid MAC TE policy.  
Android has relied heavily on MAC TE to protect important daemons and system services since its introduction in Android 5.0, as shown by the number of {\bf MAC TE allow rules}\footnote{The drastic increase of MAC allow rules can be largely attributed to the effect of Google's Project Treble~\cite{selinuxAndroid8} in Android 8, which introduced many new MAC domains due to the decomposition of the Hardware Abstraction Layer (HAL).} in Table~\ref{tab:attack_surface}.  Due to its immutable nature, the MAC TE policy provides a foundation for Android access control that other policies can modify. 

The three {\bf TE} rows (rows 2-4) of Table~\ref{tab:attack_surface} show the number of binding-IVs, write-IVs, and read-IVs for the MAC TE policy using the rules in Section~\ref{subsec:ivdef}.  We note that in counting the TE IVs we only use the MAC TE labels to identify subjects and objects, which results in coarser-grained subjects and objects than Section~\ref{subsec:background}.  Thus, the TE IV counts presented represent a lower bound.  We found this sufficient for the qualitative comparison with IV counts after TE IV validation below.  The pathname-IV count is not shown as no additional pathname-IVs are possible since there is no permission expansion allowed for the MAC TE policy.

\vspace{-0.1in}
\subsection{TE IV Validation}
\label{subsec:iv-validation}
\vspace{-0.05in}

{\em How are the number of integrity violations (IVs) reduced after TE IV Validation from those found in the TE IV Computation?}
The next four rows (rows 5-8) in Table~\ref{tab:attack_surface} show the number of IVs for the four IV types in Section~\ref{subsec:ivdef} considering after TE IV Validation ({\bf Valid}) using other Android access control policies\footnote{Note that the total IV count shown in Table~\ref{tab:sum_table} for {\em \tool IVs after Expansion} row are equal to the sum of {\em All Read-IVs} and {\em All Pathname-IVs} rows in Table~\ref{tab:attack_surface}.  The Write-IVs are a subset of the Read-IVs (i.e., all victims have read access to IVs when they have write access) and the Binding-IVs are a subset of the Pathname-IVs (i.e., victims can still access binding-IVs through luring).}.  We see that the number of TE IVs (rows 2-4) is much greater than  the number of valid IVs (rows 5-8), even accounting for the coarser subjects and objects applied in the TE IV counts\footnote{In addition, 25\% of TE IVs cannot be validated because the MAC-to-DAC mapping for some subjects is not known, see Section~\ref{subsec:our_limits}.  Although this is a large number of TE IVs, the combination of policies still reduces the Valid IV counts much more significantly.}.

Recall in Table~\ref{tab:sum_table} that the total IV counts after permission expansion are much higher across every release, showing that more testing to detect vulnerabilities is required than just testing IVs from the current policy.  However, we observed that the SEAndroid MLS policy does effectively prevent several opportunities for victim permission expansion for objects in application-private directories (e.g., {\tt /data/data}).  If MLS can be effectively applied to Android filesystems more broadly that may greatly reduce the opportunities for victim permission expansion.  

\vspace{-0.13in}
\subsection{IVs for OEM Customizations}
\label{subsec:oem}
\vspace{-0.08in}

{\em How do OEM customizations impact the Android integrity violation counts across vendors?}  To make their products more attractive, OEMs customize Android images to provide vendor-specific, value-added functionality and more attractive user interfaces. We are interested to see how OEM customization affects the number of integrity violations created when the OEMs have to customize their Android access control policies. The devices of choice are as follows: Huawei Mate9 on Android O and Android P, Xiaomi Mix2 on Android O and Android P, and Samsung Galaxy S8 on Android P. The results are shown in the right half of Table~\ref{tab:attack_surface}.

We can see heavy customization of the MAC policy. Every OEM has a significantly greater number of MAC allow rules than the Google MAC policies in the left half of Table~\ref{tab:attack_surface}.  This suggests OEMs have introduced many new domains for their own services and apps, and granted them a wide variety of MAC permissions.  The result of this customization is a significant increase MAC TE integrity violations, often more than twice as many as the associated Google Android systems.  Even more importantly, the number of integrity violations is significantly higher for the OEMs after TE IV validation (rows 5-8 in Table~\ref{tab:attack_surface}).  For example, the number of binding-IVs in Android version 9.0 systems is 32 for Google and at least 154 for the OEM Android 9.0 systems.  

\begin{table*}[t]
\centering
\begin{threeparttable}%
    \caption{Attack Operations}
    \label{tab:AOC}
     \footnotesize
\begin{tabular}{@{}c|ccc|ccccc@{}}
                                      & \textbf{Nexus 5x 7.0} & \textbf{Nexus 5x 8.0} & \textbf{Pixel3a 9.0} & \textbf{Mate9 8.0} & \textbf{Mate9 9.0} & \textbf{Mix2 8.0$^1$} & \textbf{Mix2 9.0} & \textbf{Galaxy S8 9.0} \\ \midrule
\textbf{File Attack$^2$}              & 176                & 70                   & 79                   & 597                    & 358                & 478               & 655                 & 862  \\
\textbf{Link Traversal Attack$^3$}    & 1                  & 8                    & 3                    & 169                    & 7                  & 175               & 4                   & 507    \\
\textbf{File Squat Attack$^3$}        & 52                 & 22                   & 32                   & 660                    & 443                & 248               & 154                 & 847   \\
\textbf{Pathname Attack$^3$}  & 121                & 317                  & 848                  & 734                    & 969                & 761               & 1,324                 & 2,874 
\end{tabular}%
\begin{tablenotes}\scriptsize
    \item [] Unit: Sum of operations for all (victim, object) IVs
    \item[1] This phone has significantly more files perhaps related to a higher number of pre-installed apps
    \item[2] Only for file objects
    \item[3] Only for directory objects
\end{tablenotes}
\end{threeparttable}%
\end{table*}

\vspace{-0.12in}
\subsection{IVs to Attack Operations}
\label{subsec:iv2ao}
\vspace{-0.08in}

{\em How many attack operations are really possible for the IVs found across OEM releases?}
Table~\ref{tab:sum_table} shows that not all the IVs found after permission expansion enable adversaries to launch attack operations because filesystem and/or victim subjects' program configurations may prevent attack operations, as described in Section~\ref{subsec:operations}.  

Table~\ref{tab:AOC} breaks down how many attack operations of each type are possible given the configurations that may block such operations. The number of {\em file attack} operations (adversary writes) are roughly the same as the number of read integrity violations ({\bf Valid Read-IVs}), because not many objects associated with integrity violations reside in read-only directories.  The number of {\em file squat attack} operations is the same as the number of integrity violations for directories ({\bf Valid Binding-IVs}) in Table~\ref{tab:attack_surface}.  However, the number of {\em link traversal attack} operations that are possible is fewer than the number of integrity violations because not all filesystems support symbolic links, reducing the number of directories where this attack operation applies.

The {\em luring traversal attack} operations row identifies the number of luring traversal attacks that could be performed via Binder IPC, see Section~\ref{subsec:operations}.  We can easily see that the number of operations is a lot greater than the number of binding-IVs alone ({\bf Valid Binding-IVs}), since adversaries can expand the victim's permissions for pathname-IVs ({\bf Valid Pathname-IVs}).  Recall that FileProvider usage is key to preventing luring traversal attacks (see the {\tt luring-traversal} rule in Section~\ref{subsec:operations}), where it has a non-trivial but modest impact on reducing attack operations (14-21\% across all releases).  For example, on Samsung Galaxy S8, we found that 57 out of 356 Java applications utilize FileProvider for file sharing, which meant that 3,424 pathname-IVs were only reduced to 2,874 luring-traversal operations.

\begin{table*}[t]
\centering
\begin{threeparttable}%
    \caption{Cross-Privilege Level IVs}
    \label{tab:CROSS_PRIV}
     \footnotesize
\begin{tabular}{@{}c|ccc|ccccc@{}}  
                             & \textbf{Nexus 5x 7.0} & \textbf{Nexus 5x 8.0} & \textbf{Pixel3a 9.0}  & \textbf{Mate9 8.0} & \textbf{Mate9 9.0} & \textbf{Mix2 8.0*} & \textbf{Mix2 9.0} & \textbf{Galaxy S8 9.0} \\ \midrule
\textbf{T1$^*$ $\rightarrow$ T2$^1$}$^2$ & 28      & 6                    & 17                   & 54                & 29                  & 124                & 24                 & 64                   \\
\textbf{T1 $\rightarrow$ T3} & 40                  & 22                   & 21                   & 17                & 12                  & 40                 & 25                 & 22                   \\
\textbf{T1 $\rightarrow$ T4} & 30                  & 13                   & 7                    & 14                & 8                   & 29                 & 14                 & 12                   \\
\textbf{T1 $\rightarrow$ T5} & 24                   & 9                    & 6                   & 16                & 8                   & 23                 & 8                  & 12                   \\
\textbf{T2 $\rightarrow$ T3} & 40                  & 22                   & 21                  & 20                 & 15                  & 60                 & 48                 & 92                 \\
\textbf{T2 $\rightarrow$ T4} & 30                  & 13                   & 7                   & 14                 & 8                   & 78                 & 72                 & 199                 \\
\textbf{T2 $\rightarrow$ T5} & 24                   & 9                    & 6                  & 20                 & 11                  & 34                 & 16                 & 41                  \\
\textbf{T3 $\rightarrow$ T4} & 31                  & 24                  & 16                   & 265                & 129                 & 85                 & 87                 & 124                 \\
\textbf{T3 $\rightarrow$ T5} & 68                  & 28                  & 14                   & 108                & 126                 & 42                 & 107                & 46                 \\
\textbf{T4 $\rightarrow$ T5} & 0                    & 0                    & 0                    & 0                & 0                   & 0                  & 0                  & 0                  
\end{tabular}
\begin{tablenotes}\scriptsize
    \item[*] T1(untrusted/isolated app), T2(priv/platform app) T3(services), T4(system app, system service), T5(root service)
    \item[1] For adversary at lower level (T1) and victim at higher level (T2)
    \item[2] Unit: Sum of binding and file IVs (no pathname-IVs included)
\end{tablenotes}
\end{threeparttable}
\end{table*}

\vspace{-0.12in}
\subsection{Cross-Privilege Level IVs}
\label{subsec:cross}
\vspace{-0.08in}
{\em How are integrity violations distributed across Android privilege levels?}  
The IV distribution is important because it indicates how victims at each privilege level could be attacked and how adversaries at any privilege level could compose attacks to reach other privilege levels. Table~\ref{tab:CROSS_PRIV} shows the counts of file and binding integrity violations between each pair of privilege levels we evaluated.  We do not include pathname-IVs in this table to assess attack paths without luring.  

Google's 8.0 and 9.0 releases have a modest number cross-privilege level IVs. This confirms our hypothesis that Google's access control policies are the closest to best practice.  However, on the OEM side, it can be a completely different story. Other than the Mate9 9.0, the IVs between each privilege level pair can be significant, meaning that even without luring, releases may be vulnerable in a variety of ways.  

\vspace{-0.12in}
\subsection{Vulnerability Case Studies}
\label{subsec:case}
\vspace{-0.03in}

{\em What kind of vulnerabilities may be discovered from attack operations?}
We found two previously unknown vulnerabilities manually using the attack operations computed by \tool.  

\paragraph{Samsung Resetreason}
We also found a new binding vulnerability in the Samsung Galaxy S8 system using the Android 9.0 release.  Samsung includes a privileged service called {\em resetreason} that logs the reason why the phone has had to reset into the file {\tt power\_off\_reset\_reason.txt} in the directory {\tt /data/log}.  However, any process that runs with the {\tt AID\_LOG} group has write permission to that file, so such processes can replace the file with a symbolic link to any file accessible to {\em resetreason} to launch a link traversal attack.  While only signed apps may be granted the Android permission ({\tt READ\_LOGS}) associated with the {\tt AID\_LOG} DAC group, vendors deploy several signed apps on their systems, and some signed apps have had reported vulnerabilities, such as the {\tt adb} app~\cite{adbport}.
{\tt resetreason} has access to several integrity-critical resources, and we have confirmed that we can redirect {\tt resetreason} to write files in the encrypted filesystem directory.  
Previous work demonstrated the importance of attacks on the encrypted filesystem from the system's radio service~\cite{tian18usenix}. This vulnerability has been confirmed by Samsung and assigned CVE ID CVE-2020-13833.

\paragraph{Xiaomi and Huawei Thememanager}
We discovered multiple unreported vulnerabilities in the Xiaomi and Huawei systems.  We describe one example here.  
These systems include a variety of value-added services, where one is the Thememanager, which allows users to customize the user interface of their devices. However, the Xiaomi 
access control policies are configured such that untrusted apps can write to the file {\tt /data/data/com.android.thememanager/cache}, which is used by the Thememanager for storing content that the Thememanager may use in configuring the display.  We verify on Xiaomi 8.0 that arbitrary modifications to this file do crash the privileged Thememanager process and in some cases impact the GUI without crashing the Thememanager.  A finely-crafted modification could perhaps exploit the Thememanager service.  We found four other similar vulnerabilities in the Xiaomi 8.0 release for writeable cache files. 

We found that Huawei on both the 8.0 and 9.0 releases has a similar vulnerability for the theme cache files as well, but exploitation requires adversaries to compromise a privileged application with {\tt media\_rw} permission.  However, privileged applications have been found to be flawed in several instances, see Section~\ref{subsec:limits}.  


\vspace{-0.12in}
\subsection{Performance}
\label{subsec:perf}
\vspace{-0.05in}

We measure the performance of \tool for the eight Android releases.  The overhead was measured on a PC running an AMD Ryzen 7 3700X (8 core, 16 thread) with 16GB of RAM and an RTX 2080 Super GPU using Ubuntu 18.04.  \tool IVs are found in two steps as described in Section~\ref{sec:impl}: TE IV computation and TE IV validation. We find that the performance of TE IV computation has a linear relationship to the SEAndroid policy size. The TE IV validation stage's performance is proportional to the number of IVs found in TE IV computation, but that impact can be reduced because validation can be parallelized.   

\begin{figure}[t]
\centering
\includegraphics[width=8cm]{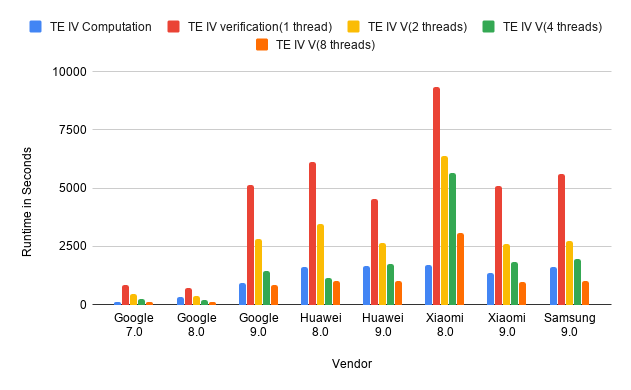}
\vspace{-1.5em}
\caption{{\tool Performance}}
\label{fig:perf}
\end{figure}

Figure~\ref{fig:perf} shows the performance overhead of these two stages\footnote{The cost of computing attack operations is negligible and included in the TE IV validation.} for the eight releases.  We evaluate the performance of the TE IV validation for one to eight threads.  With a multi-core CPU, parallelization does produce significant performance improvement. We also point out that we found it quite expensive to compute all the authorized data flows for these releases.  On the other hand, with a proper threat model to prune cases, \tool is able to identify integrity violations in a reasonable amount of time.

\vspace{-0.12in}
\section{Discussion}
\label{sec:discussion}
\vspace{-0.05in}

In this section, we review limitations in the \tool approach and examine the implications of a recently proposed Android defense called {\em scoped storage}.

\vspace{-0.12in}
\subsection{Limitations}
\label{subsec:our_limits}
\vspace{-0.05in}

We identify three limitations of \tool: (1) \tool relies on rooted phone to collect filesystem data; (2) we cannot always determine the mapping between MAC labels and their corresponding DAC UIDs; (3) \tool cannot confirm vulnerabilities from attack operations automatically.  

Without rooting the phone, we cannot gather DAC information from privileged directories like {\tt /system}.  Recent research developed a tool called BigMAC~\cite{BigMAC}, which extracts accurate DAC configuration data from these privileged directories (\~95\%).  We will include BigMAC into our data collection in future iterations.  Together with the data collected from an unrooted phone, see Appendix~\ref{subsec:collect}, we should be able to recover a nearly complete snapshot of the filesystem.  We will explore methods to achieve complete recovery in future work.

Another limitation of \tool is that finding the MAC-to-DAC mapping of subjects requires running a process for each MAC label to collect its DAC UIDs/groups\footnote{Recall that we leverage the finding of Chen {\em et al.}~\cite{chen17acsac} that the MAC-to-DAC mapping for Android systems is one-to-one.}.  Thus, it is possible no process was launched for some MAC labels.  Currently, if either no adversary or victim for a computed TE IV is mapped to a complete subject, we skipped the IV validation stage for that TE IV. Currently, about 25\% of the TE IVs do not go through validation.  Runtime support could detect when a process is run for an unmapped MAC label to collect the mapping.

Finally, \tool lacks a systematic way to test the victims for vulnerabilities to the attack operations found.  The problem is that we need to know when a victim uses a file, binding, or IPC that is associated with an attack operation.  A passive runtime tool~\cite{sting} was developed to monitor processes for use of bindings associated with attack operations (i.e., file squatting and link traversal).  However, this tool only used the available DAC policies to determine whether an attack operation would be possible, and did not test for other attack operations.  \tool's more accurate computation of attack operations should improve the effectiveness of such an approach.  To test for luring traversals, one must develop methods to detect at-risk IPCs rather than file accesses.  The Jigsaw system~\cite{jigsaw14usenix} provides a method for identifying system calls that may receive input that could enable luring traversals, but it does not identify the scope of targets to which luring may occur.   \tool identifies a full scope of luring targets using victim permission expansion, so we will explore the use of \tool to generate test cases for the system calls identified by Jigsaw.

\vspace{-0.1in}
\subsection{Scoped Storage}
\label{subsec:scoped}
\vspace{-0.05in}

Polyscope identifies several integrity violations located in external storage, so we discuss the potential impact  on \tool of a defense introduced in Android 10 that offers some protections for external storage, called {\em scoped storage}~\cite{scopedstorage}. Scoped storage aims to remove two coarse-grained and problematic Android Permissions that grant applications read/write permissions to external storage. Under fully enforced scoped storage, applications can only access their private folder in external storage, and file sharing should only be performed with designated file sharing APIs like {\tt FileProvider}. However, to support application functionality and ease the transition, Android 10 does not fully enforce scoped storage. Applications can still store files in shared external storage locations and have the option to opt-out of scoped storage by requesting for legacy storage in the manifest file. Therefore, IVs identified by Polyscope are still potential targets in Android 10. For the upcoming Android 11, current documentation suggests that scoped storage will be fully enforced, but a functionality named \emph{All File Access}\cite{AFA} can still grant applications read-write permission to files stored on external storage. This could potentially open up new attack surfaces and once the technical details of scoped storage are finalized for Android 11, we will include the changes in \tool.  Finally, we note that scoped storage has no impact on IVs outside of external storage, including the vulnerable IVs identified in Section~\ref{subsec:case}.

\vspace{-0.1in}
\section{Related Work}
\label{sec:relwork}
\vspace{-0.07in}


Researchers have long known about the three types of integrity violations listed in Section~\ref{sec:threat}, but have found it difficult to prevent programs from falling victim to such threats.  A variety of mechanisms have been proposed to prevent attacks during name resolution, including defenses for binding and pathname vulnerabilities.  These defenses have often been focused on TOCTTOU attacks~\cite{mcphee74,bishop-dilger}.  Some defenses are implemented in the program or as library extensions~\cite{raceguard,27,dean-hu,tsafir} and some as kernel extensions~\cite{chapin_tocttou,28,venema_ndss_2010,openwall,ty-race,35}, but the methods overlap, where some enforce invariants on file access~\cite{raceguard,chapin_tocttou,35,27,28,ty-race}, some enforce namespace invariants~\cite{venema_ndss_2010,openwall}, and some aim for ``safe'' access methods~\cite{dean-hu,tsafir}.  In general, all program defenses have been limited because they lack insight into the changing system and all system defenses are limited because they lack side-information about the intent of the program~\cite{johnson-tocttou}.  


The main defense for preventing filesystem vulnerabilities is access control.  If the access control policies prevent an adversary from accessing the filesystem resources that enable attack operations, then the system is free of associated vulnerabilities.  However, the discretionary access control (DAC) policies commonly used do not enable prediction of whether a subject may obtain an unauthorized permission~\cite{hru76}, so techniques to restrict DAC~\cite{take-grant,tam,tidswell00ccs} and apply mandatory access control (MAC) enforcement~\cite{blp76,b77} were then explored, culminating in the adoption of MAC enforcement in some systems, such as Linux Security Modules~\cite{wcs+02} employed in SELinux~\cite{selinux} and AppArmor~\cite{apparmor}.  Researchers than proposed MAC enforcement for Android systems~\cite{xie09srds,bugiel12ndss}, but a version of SELinux~\cite{selinux} targeting Android was developed, called SEAndroid~\cite{smalley2013security} was adopted.  However, the attack operations we find in this paper abuse available MAC permissions.  While a techniques have been developed to limit processes to their full permissions on individual system calls~\cite{cw-lite,process-firewall}, such techniques need policy analysis to determine the policies to enforce.



Researchers have proposed using access control policy analysis to identify misconfigurations that may lead to vulnerabilities~\cite{jaeger02sacmat,setools}, but traditionally, access control policy analysis methods only reason about one policy, such as a mandatory access control (MAC) policy~\cite{setools,jaeger03usenix,chen09ndss,integrity-wall} or an Android permission policy~\cite{enck09ccs,wae+17,Wang2015}. However, based on the research challenges above, we must consider the combination of the access control policies employed on the system to compute attack operations accurately.  Chen {\em et al.}~\cite{chen17acsac} were the first work that we are aware of to combine MAC and DAC policies in access control policy analysis. Hernandez {et al.}~\cite{BigMAC} further extended their analysis to include MAC, DAC and Linux capabilities.  However, both of these techniques compute data flows, which are much more numerous than integrity violations.  Chen {\em et al.} look for data flows that may lead to sensitive data leakage directly rather than attack operations that may enable such leakage as \tool does.

\vspace{-0.13in}
\section{Conclusions}
\label{sec:conc}
\vspace{-0.12in}

Android uses a combination of filesystem access control mechanisms to assure its platform integrity.
This paper has proposed \tool, a policy analysis tool that reasons over Android's mandatory (SEAndroid) and discretionary (UNIX permissions) access control policies, in addition to the other mechanisms (e.g., Android permissions) that influence file access control.
\tool is novel in its ability to reason about permission expansion, which lies at the intersection of mandatory and discretionary policy.
We applied \tool to three different Google Android releases and five different OEM Android releases, characterizing the potential for file-based attacks such as file squatting, link traversal, and luring traversal.
In doing so, we identified two new vulnerabilities in OEM Android releases and opportunities to direct further automated testing.
Our results suggest that the access control policy changes introduced by OEMs do not sufficiently address integrity violations for their feature additions.

\bibliographystyle{plain}
\bibliography{ref,firewall_long}


\appendix

\vspace{-0.1in}
\section{Additional Background}
\label{app:background}
\vspace{-0.07in}

In this section, we provide details on how \tool collects the relevant access control information.

\subsection{Access Control Data Collection}
\label{subsec:collect}

\paragraph{MAC Data} To obtain MAC data, \tool first pulls the SEAndroid policy binary file from the Android root directory with command "\texttt{adb pull sepolicy}". With the SELinux policy binary in hand, we extract the allow rules with "\texttt{sesearch -A sepolicy}". Then, in order to parse the SELinux attributes, we pull the attribute mapping with "\texttt{seinfo -a -x sepolicy}".

\paragraph{DAC Data} To obtain DAC permissions for all files on an Android system, \tool executes "\texttt{adb shell ls -lRZ}" from the root directory. Note that the phone must be rooted to obtain the full list of file permissions, so we use a boot time root technique to gain root. \tool collects the file permission data shown in Table~\ref{DAC data}.
The data in Table~\ref{DAC data} indicates: a file \texttt{authtokcont} under the directory \texttt{/efs} has \texttt{read, write} permissions for its owner and group members. Its owner and group UID are both \texttt{radio}, and its MAC security label is \texttt{efs\_file}.
\begin{table}[h]
\caption{File DAC data sample}
\label{DAC data}
\centering
\footnotesize
\begin{tabular}{ccccc}
\textbf{File}    & \textbf{DAC perms} & \textbf{User} & \textbf{Group} & \textbf{MAC security label} \\ \midrule
\texttt{authtokcont} & \verb/-rw-rw-r--/         & \texttt{radio}         & \texttt{radio}          & \texttt{efs\_file}                   \\ 
\end{tabular}%
\vspace{-0.2in}
\end{table}

\paragraph{Process Information} \tool obtains process access control information by executing the command "\texttt{adb shell ps -A -o label,user,group,COMMAND}", which provides a mapping from a DAC user ID to MAC label for running processes. One data sample is shown in Table~\ref{process data}.  This  entry shows that process \texttt{init} has security label of \texttt{u:r:init:s0}, UID of \texttt{root}, GID of \texttt{root}, was spawned by command \texttt{/init}, and PID of \texttt{1}. However, the process list collection does not provide the full information on DAC supplementary groups, as we described in Section~\ref{subsec:expand}. 
In the case of Android system services, these extra groups are defined in the \texttt{init.rc} file, which can be parsed statically. For apps, \tool uses a shell script to obtain process DAC group information stored in \texttt{/proc}.

\begin{table}[h]
\caption{Process Data}
\label{process data}
\centering
\footnotesize
\begin{tabular}{ccccc}
\textbf{Security label} & \textbf{UID} & \textbf{Group}  & \textbf{Command} & \textbf{PID} \\ \midrule
\texttt{u:r:init:s0}                    & \texttt{root}         & \texttt{root}                  & \texttt{/init} & \texttt{1}    \\
\end{tabular}%
\vspace{-0.2in}
\end{table}

\paragraph{Android Permission Data} To obtain Android Permissions' mappings to DAC groups, \tool parses \texttt{/etc/platform.xml} from the Android device. Next, we need to separate the signature Android Permissions from the non-signature Android Permissions, which are available via the Android package manager (PM), as the non-signature permissions may be applied by an app.  \tool uses the non-signature permissions to compute DAC expansion for adversaries.

\paragraph{Filesystem and FileProvider} To determine whether attack operations are blocked, \tool needs to examine the filesystem configuration and the application package.  First, \tool obtains filesystem configurations by running "\texttt{adb shell mount}", which will return list of filesystem mount configuration. We identify the directories mounted with the {\texttt ro} or the {\texttt nosymlink} flags and mark them as read-only and prohibiting symlinks, respectively.  Second, for the application package, we want to determine if the application uses the FileProvider class to protect itself from luring.  \tool first queries the PackageManager service for a full list of apk files on the system.  Next, \tool collects all the apk files found and performs code inspection with Google's new ClassyShark tool~\cite{ClassyShark} to identify the presence of the FileProvider class.

\end{document}